\documentclass[10pt]{article}

\usepackage{amsxtra,amssymb,amsthm,amsmath,latexsym}

\usepackage{graphicx}

\newtheorem{thm}{Theorem}

\newtheorem{lem}{Lemma}

 \newcommand{\thmref}[1]{Theorem~\ref{#1}}

\newcommand{\R}{{\mathbb R}}
\newcommand{\C}{{\mathbb C}}

\newcommand{\ep}{\epsilon}

\newcommand{\dl}{{\delta}}
\newcommand{\bee}{\begin{equation*}}
\newcommand{\eee}{\end{equation*}}
\newcommand{\be}{\begin{equation}}
\newcommand{\ee}{\end{equation}}
\newcommand{\pn}{\par\noindent}
\title{Inverse scattering with non-overdetermined data}
\author{A G Ramm\\
\small Department of Mathematics\\[-0.8ex]
\small Kansas State University, Manhattan, KS 66506-2602, USA\\[-0.8ex]
\small \texttt{ramm@math.ksu.edu}\\
}

\begin{document} \date{} \maketitle \begin{abstract} 
Let $A(\beta,\alpha,k)$ be the scattering amplitude corresponding to a
real-valued potential which vanishes outside of a bounded domain 
$D\subset \R^3$.
The unit vector $\alpha$ is the direction of the incident plane
wave, the unit vector $\beta$ is the direction of the scattered wave,
$k>0$ is the wave number. 
The governing
equation for the waves is $[\nabla^2+k^2-q(x)]u=0$ in $\R^3$. 

For a suitable class of potentials it is proved  that if
$A_{q_1}(-\beta,\beta,k)=A_{q_2}(-\beta,\beta,k)$ $\forall   
\beta\in S^2,$ $\forall k\in (k_0,k_1),$ and $q_1,$ $q_2\in
M$, then $q_1=q_2$. This is a uniqueness theorem for the solution to 
the inverse scattering problem with backscattering data.

It is also proved for this class of potentials that if
$A_{q_1}(\beta,\alpha_0,k)=A_{q_2}(\beta,\alpha_0,k)$ $\forall
\beta\in S^2_1,$ $\forall k\in (k_0,k_1),$ and $q_1,$ $q_2\in M$,
then $q_1=q_2$. 

Here $S^2_1$ is an arbitrarily small open subset of
$S^2$, and $|k_0-k_1|>0$ is arbitrarily small.

\end{abstract}

\pn{
 }

\section{Introduction}
Consider the scattering problem:
\be\label{e1} Lu:=[\nabla^2+k^2-q(x)]u=0\quad in\quad \R^3,\quad
k=const>0, \ee 
\be\label{e2} u=e^{ik\alpha\cdot
x}+A(\beta,\alpha,k)\frac{e^{ikr}}{r}+o\left(\frac{1}{r}\right),\quad
r:=|x|\to \infty,\quad \beta=\frac{x}{r},\quad \alpha\in S^2, \ee
where $S^2$ is the unit sphere in $\R^3$, and
$A(\beta,\alpha,k)=A_q(\beta,\alpha,k)$ is the scattering amplitude
corresponding to the potential $q(x),$ $\alpha$ is the direction of the
incident plane wave, $\beta$ is a direction of the scattered wave,
and $k^2$ is the energy.

 Let us assume that $q$ is a
real-valued compactly supported function,  
$$q\in
M:=W^{\ell,1}_0(D),\quad \ell>2,$$ $D\subset \R^3$ is a bounded domain,
and $W^{\ell,1}_0(D)$ is the Sobolev space,
it is the closure of $C^\infty_0(D)$ in the norm of the Sobolev space
$W^{\ell,1}(D)$. This space consists of the functions whose derivatives up 
to the order $\ell$ are absolutely integrable in $D$.

The inverse scattering problems, we are studying in this paper, are:

{\it IP1: Do the backscattering data $A(-\beta,\beta,k)$ known $\forall k>0$,
$\forall
\beta\in S^2,$ determine $q\in M$ uniquely?}

{\it IP2: Do the data $A_q(\beta,k):=A(\beta,\alpha_0,k)$
known $\forall k>0$, $\forall \beta\in S^2,$ determine $q\in M$
uniquely?}

We give a positive answer to these questions. Theorem 1 (see below) is
our basic result.

These inverse problems have been open for many decades (see, 
e.g., \cite{R470}). They are a
part of the general question in physics: does the $S$-matrix
determine the Hamiltonian uniquely?\\
It was known that the data $A(\beta,\alpha,k)$ $\forall
\alpha,\beta\in S^2$, $\forall k>0$, determine $q(x)\in
C^1(\R^3)\cap C(\R^3,(1+|x|)^\gamma,\,\gamma>3)$ uniquely. Here
$\|q\|_{C(\R^3,(1+|x|)^\gamma)}=\sup_{x\in\R^3}\{(1+|x|)^\gamma|q(x)|\}$,
and the datum $A(\beta,\alpha,k)$ is a function of $5$ variables
(two unit vectors $\beta,\alpha \in S^2$ and a scalar $k>0$), while 
the potential $q$
is a function of $3$ variables, $(x_1, x_2, x_3)$.
We are not stating this old result with minimal assumptions on the 
class of potentials.
 
The author proved (see \cite{R228}- \cite{R470})
that the data $A_q(\beta,\alpha):=A_q(\beta,\alpha,k)$, known
$\forall \alpha\in S^2_1$, $\forall \beta\in S^2_2$ and a fixed
$k=k_0>0$, determine $q\in Q_a$ uniquely. Here $S_j^2,$ $j=1,2$, are
arbitrary small open subsets of $S^2$ (solid angles),  and
$$Q_a:=\{q:q=\overline{q},q=0\quad if\quad |x|>a,\quad q\in
L^2(B_a)\}, \quad B_a:=\{x:\ |x|\leq a\},$$
$a>0$ is an arbitrary large fixed number. 
In this uniqueness theorem the datum $A_q(\beta,\alpha)$ is a function of four
variables (two unit vectors $\alpha, \beta \in S^2$) and the potential $q$ is
a function of three variables $(x_1,x_2,x_3)$. Therefore, this inverse
problem is also overdetermined.

It is natural to assume that $q$ has compact support
in  a study of the inverse scattering problem, because in practice
the data are always noisy, and from noisy data it is {\it in principle
impossible} to determine the rate of decay of a potential $q(x),$
such that $|q(x)|\leq c(1+|x|)^{-\gamma},$ $\gamma>3$, for all
sufficiently large $|x|$. Indeed, the contribution of the "tail" of
$q$, that is, of the function  $q_R:=q_R(x)$, \bee q_R(x):=\left\{
                                                        \begin{array}{ll}
                                                0, & \hbox{$|x|\leq R$,} \\
                                                          q(x), & \hbox{$|x|>R$,}
                                                        \end{array}
                                                      \right.\eee
to the scattering amplitude cannot
be distinguished from the contribution of the noise if $R$ is
sufficiently large. For example, if the noisy data are $A_q^{(\dl)}(\beta, \alpha, k)$,
$$\sup_{\beta,\alpha\in
S^2}|A_q^{(\dl)}(\beta,\alpha,k)-A_q(\beta,\alpha,k)|<\dl,$$
then one can prove  that  the contribution of $q_R$ to $A_q$ is
$O\left(\frac{1}{R^{\gamma-3}}\right).$ Thus, this contribution is of the
order of the noise level $\dl$ if $R=O(\dl^{1/(3-\gamma)})$, $\gamma>3$.
This yields an estimate of the "radius of compactness" of the potential
$q$ given the noise level $\delta$ and the exponent $\gamma>3$, which
describes the rate of decay of the potential.
 
There were no results concerning the uniqueness of the solution to
the inverse scattering problems IP1 and IP2 with  the non-overdetermined 
backscattering data $A(-\beta,\beta,k)$ $\forall \beta\in S^2$, $\forall 
k>0$,
or with the  non-overdetermined data $A(\beta,\alpha_0,k)$
$\forall \beta\in S^2$, $\forall k>0$, $\alpha=\alpha_0$ being fixed.

The main result of this paper is:
\begin{thm}\label{thm1}

1) If $A_{q_1}(-\beta,\beta,k)=A_{q_2}(-\beta,\beta,k)$
$\forall \beta\in S^2,$ $\forall k>0$ and  $q_j\in M$, $j=1,2,$
then $q_1=q_2.$

2) If $A_{q_1}(\beta,\alpha_0,k)=A_{q_2}(\beta,\alpha_0,k)$
$\forall \beta\in S^2,$ $\forall k>0$, $\alpha_0\in S^2$ is fixed,
and $q_j\in M,$ $j=1,2,$ then $q_1=q_2.$

\end{thm}
{\it Remark 1.} \thmref{thm1} remains valid if the data are given $\forall
\beta\in S_1^2$, $\forall k\in (k_0,k_1)$, $0<k_0<k_1$, where $S^2$
and $|k_1-k_0|>0$ is arbitrarily small.\\
Indeed, if $q\in M$, or, more generally, if
$q$ is compactly supported, supp $q\subset B_a$, and $q\in L^2(B_a)$,
then the author has proved (see \cite{R470} and \cite{R190}), that
$A(\beta,\alpha,k)$ is a
restriction to $(0,\infty)$ of a meromorphic in $\C$ function of $k$
and a restriction to $S^2\times S^2$ of a  function analytic on the variety
$\mathcal{M}\times\mathcal{M}$, $\mathcal{M}:=\{\theta:\, \theta \in 
\C^3,\, \theta\cdot\theta=1\}$, where
$\theta\cdot\theta:=\sum_{j=1}^3\theta_j^2.$
Therefore, if
$A(\beta,\alpha_0,k)$ is known on $S_1^2\times (k_0,k_1)$ then it is
uniquely determined on $S^2\times(0,\infty)$ by  analytic
continuation.\\
The algebraic variety
$\mathcal{M}$ is a non-compact algebraic variety in  $\C^3$.

{\it Remark 2.} The main idea of the proof of
Theorem 1 is to establish completeness of the set of products of
the scattering solutions in a class $M$ of potentials.
This is a version of Property C, introduced and applied by the author
to many inverse problems (see \cite{R220}, \cite{R252}, \cite{R262},  \cite{R470}).

\section{Proofs}
The following lemma is crucial for the proof of both statements of
Theorem 1.
\begin{lem}(\cite[p.262]{R470})\label{lem1} If $p(x):=q_1(x)-q_2(x)$, then
\be\label{e4}
-4\pi[A_{q_1}(\beta, \alpha, k)-A_{q_2}(\beta, \alpha, k)]=
\int_Dp(x)u_1(x, \alpha, k)u_2(x, -\beta, k)dx.
\ee
\end{lem}

In \eqref{e4} $u_j$ are the scattering solutions, that is, solutions to 
\eqref{e1}-\eqref{e2} with $q=q_j$, or, equivalently, solutions to the 
integral equation:
\be\label{e5}
u_j(x, \alpha, k)=e^{ik \alpha \cdot x} -\int_D g(x,y,k) q_j(y) 
u_j(y, \alpha, k)dy, \quad  g(x,y,k):=\frac {e^{ik|x-y|}}
{4\pi |x-y|}.
\ee
Let $v_j:=e^{-ik \alpha \cdot x}u_j$. Then 
\be\label{e6}
u_j=e^{ik \alpha \cdot x} [1+\epsilon_j],\quad \epsilon_j:= -\int_D G(x,y,k) q_j(y)v_j(y, \alpha, k)dy, 
\ee
where 
$$ G(x,y,k):=g(x,y,k)e^{-ik\alpha \cdot (x-y)}.$$
The function $v_j$ solves the integral equation
\be\label{e6'}
v_j=1-B_jv_j, \qquad B_jv_j:=-\int_D G(x,y,k) q_j(y)v_j(y, \alpha, k)dy,
\ee
and $B_jv_j=\ep_j$.

If $A_{q_1}=A_{q_2}$ $\forall \beta\in
S^2,$ $\forall k>0$, and $\beta =-\alpha$, then \eqref{e4} yields the following
{\it orthogonality relation}: 
\be\label{e7}
\int_Dp(x)u_1(x,\beta, k)u_2(x,\beta,k)dx=0,\quad \forall
\beta\in S^2,\quad \forall k>0, 
\ee
where
$$ p(x)=q_1(x)-q_2(x).$$
The IP2 is treated similarly.

The orthogonality relation \eqref{e7} can be written as
\be\label{e8}
\int_Dp(x)e^{2ik\beta \cdot x}[1+\epsilon (x,\beta, k)]dx=0,\quad \forall
\beta\in S^2,\quad \forall k>0, \quad  
\epsilon:=\epsilon_1+\epsilon_2 +\epsilon_1 \epsilon_2.
\ee
The relation \eqref{e8}  holds for $\Im k\geq 0,\, k\neq i\kappa_{m,j}$,
where $i\kappa_{m,j}$, $1\leq m\leq m_j$, $j=1,2,$ are the numbers at which
the operator $I+B_j$ is not injective. There are finitely many such 
numbers in the upper half complex plane if $q_j\in M$. The numbers  $\kappa_{m,j}>0$,
$-\kappa_{m,j}^2$ are the negative eigenvalues of the Schroedinger 
operator $L_j$ in $L^2(\R^3)$, where $L_j$ is the operator in 
 \eqref{e1} with $q=q_j$. 

In what follows we write $\epsilon$ meaning $\epsilon_j$ for
$j=1,2,$ or $\epsilon$, defined
in  \eqref{e8}. Also, we write $\kappa_{m}$ in place of $\kappa_{m,j}$.
This will not cause any confusion.

Since
$q$ is compactly supported, the scattering solution $u(x,\alpha,k)$ is analytic in the
region Im $k\geq 0$, except, possibly, for a finite number of poles
$k_m=i \kappa_m$, $\kappa_m>0$, $\kappa_m<\kappa_{m+1}$, 
$1\leq m \leq m_0< \infty$,  where $m_0<\infty$ is a positive integer.
Therefore, $u(x,\alpha,k)$
and $\epsilon (x,\alpha, k)$ are analytic in the region 
$\Im k\geq 0$, $k\neq k_m$,  $1\leq m \leq m_0$.
Let $\eta_0>0$ be chosen so that
$\eta_0>\max_{m}\kappa_m$.

The  orthogonality relation  \eqref{e8} for $q_j\in M$ holds 
in the region $\Im k\geq 0$, $k\neq i\kappa_m$, and the integrand
in  \eqref{e8} is analytic with respect to $k$ in this region.

We want to derive from \eqref{e8} that $p(x)=0$.

Write the orthogonality relation  \eqref{e8} as:
\be\label{e9}
\tilde{p}(2k\beta)+(2\pi)^{-3} \tilde{p}\star \tilde{\epsilon}=0,
\ee
where the $\star$ denotes convolution,
\be\label{e10}
\tilde{p}(\xi):=\int_{\R^3}e^{i\xi \cdot x}p(x)dx, \qquad  
\tilde{p}\star \tilde{\epsilon}:=
\int_{\R^3} \tilde{p}(\xi -\nu) \tilde{\epsilon}(\nu) d\nu,
\ee
and in  \eqref{e9} $\tilde{p}\star \tilde{\epsilon}$ is calculated
at $\xi=2k\beta$.

Equation  \eqref{e9} has only the trivial solution $\tilde{p}=0$ provided that 
\be\label{e11}
(2\pi)^{-3}||\tilde{\epsilon} (\xi,\beta, k)||_1<b<1, 
\ee
where
$$ ||\tilde{\epsilon}||_1=\int_{\R^3}|\tilde{\epsilon}(\xi, 
\beta , k)|d\xi.$$
Indeed,
\be\label{e12}
\max_{k\geq 0, \beta\in S^2}|\tilde{p}(2k\beta)|\leq \max_{k\geq 0, \beta\in S^2, \nu\in \R^3}|\tilde{p}(2k\beta-\nu)|
\cdot ||\tilde{\epsilon}||_1<\max_{k\geq 0, \beta\in S^2}|\tilde{p}(2k\beta)|,
\ee
where we have taken into account that the sets 
$$\{2k\beta\}_{\forall k\geq 0, \forall \beta\in S^2}$$ 
and
$$\{2k\beta- \nu\}_{\forall k\geq 0, \forall \beta\in S^2, \forall 	 \nu\in \R^3}$$ 
are the same.

Inequalities  \eqref{e11} and \eqref{e12} imply
 $$\tilde{p}(2k\beta)=0 \quad \forall k>0, \forall \beta\in S^2.$$ 
If $\tilde{p}(2k\beta)=0$ $\forall k>0$, $\forall 
\beta\in S^2$,
then $\tilde{p}=0$, and, by the injectivity of the Fourier transform, one concludes 
that $p=0$.

Since $p$ is compactly supported, the
function $\tilde{p}$ is entire function of $\xi$. Consequently, if one 
proves that
 $\tilde{p}(2(k+i\eta)\beta)=0$ $\forall k>0$, $\forall
\beta\in S^2$, and for $\eta>\eta_0>0$,
then $\tilde{p}=0$ by analytic continuation, and, consequently,  $p=0$. 
This observation is used below.

Thus, to prove the first claim of Theorem 1, it is sufficient 
to establish inequality  \eqref{e11}.

{\it However, \eqref{e11} with $k>0$ does not hold} because 
the function $\frac 1 {|\xi|^2 -2k\beta \cdot \xi}$
(see formula  \eqref{e16} below) is not absolutely integrable if $k>0$.

The idea, that makes the proof work, is to replace $k>0$ with $k+i\eta$, 
where $\eta>\eta_0>0$
is sufficiently large. The orthogonality relation  \eqref{e7} 
remains valid after such a replacement because of the analyticity
of $\epsilon=\epsilon (x,\beta, k)$ with respect to $k$ in the 
region $\Im k>\eta_{0}$. Equation 
 \eqref{e8} holds with $k+i\eta$ replacing  $k$.

The argument, given in  \eqref{e12},  remains valid after this replacement 
because 
$$\mu:=\max_{k>0,\eta\in(\eta_0,\eta_1), \beta\in 
S^2}|\tilde{p}(2(k+i\eta)\beta)|\geq
c \max_{\xi\in \R^3}|\tilde{p}(\xi)|:=c\mu_1,$$
where $c>0$ is a constant and $\eta_1>\eta_0$ is a sufficiently large
number, which is assumed finite in order to have $\mu<\infty$.

Therefore, \eqref{e9} with $k+i\eta$ replacing $k$  yields:
$$\mu\leq \max_{k>0,\eta\in(\eta_0,\eta_1),\beta\in 
S^2}\int_{\R^3}|\tilde{\epsilon}
(2(k+i\eta)\beta -\xi)|d\xi\,\, \mu_1< \mu,$$
and, consequently, $\mu=0$ and $p(x)=0$,  
provided that an analog of \eqref{e11} holds:
$$\max_{k>0,\eta\in(\eta_0,\eta_1),\beta\in 
S^2}\int_{\R^3}|\tilde{\epsilon}
(2(k+i\eta)\beta -\xi)|d\xi<b(\eta),$$
where 
$$\lim_{\eta\to +\infty}b(\eta)=0,$$ so that 
$$cb(\eta)<1,   \qquad \eta>\eta_0,$$ 
for sufficiently large $\eta>\eta_0$. 

We refer to this inequality also as \eqref{e11}, and prove
that this inequality holds if $\eta$ is sufficiently large
(see \eqref{e18} below, from which it follows that  
$$b(\eta)=O(|\eta|^{-1}) \qquad \eta\to +\infty.$$

Let us check that 
$$\mu\geq c\mu_1.$$
 This inequality will be established if one proves that
$$\mu=\sup_{\beta \in S^2, k>0, 
\eta\in(\eta_0,\eta_1)}|\tilde{p}((k+i\eta)\beta)|\geq
c\int_D|p(x)|dx,$$
because 
$$\sup_{\xi \in \R^3}|\tilde{p}(\xi)|\leq \int_D|p(x)|dx.$$
One has 
$$\mu\geq \sup_{\beta \in S^2,\eta\in(\eta_0,\eta_1)}|\int_De^{-2\eta 
\beta 
\cdot x}p(x)dx|= \sup_{\beta \in S^2,\eta\in(\eta_0,\eta_1)}|W|,$$
where 
$$W:=\int_De^{-2\eta \beta\cdot x}p(x)dx.$$
Let us prove that
$$ \sup_{\beta \in S^2,\eta\in(\eta_0,\eta_1)}|W|\geq c\int_D|p(x)|dx.$$
If this inequality is established, then the proof of the
inequality $\mu\geq c\mu_1$ is complete.

We may assume that $p\not\equiv 0$, because otherwise there is nothing
to prove. If $p\not\equiv 0$, then $W\not\equiv 0$. The function 
$W$ is an entire function of the vector $\eta \beta$, considered as a 
vector in $\C^3$. The function $\sup_{\beta \in S^2}|W|$ tends to $\infty$ 
as $\eta\to +\infty$ (see \cite{L} for the growth rates of
entire functions of exponential type). Therefore inequality
$ \sup_{\beta \in S^2,\eta\in(\eta_0,\eta_1)}|W|\geq c\int_D|p(x)|dx$ 
holds, and
inequality  $\mu\geq c\mu_1$ is established.

If inequality  \eqref{e11} is proved for $k+i\eta$ replacing $k$, then
the argument, similar to the one, given in   \eqref{e12}, yields
$\tilde{p}(2(k+i\eta)\beta)=0$ for all $k>0$, $\beta\in S^2$, 
and $\eta>\eta_{0}$.
By the analytic continuation this implies $\tilde{p}(\xi)=0$ for 
all $\xi$, so $p(x)=0$. 

The first claim of  Theorem 1 is therefore proved as soon as estimate 
 \eqref{e11} is proved with $k+i\eta$ replacing $k$.

Let us now establish  inequality  \eqref{e11} with  $k+i\eta$ replacing  $k$.

Note that 
$$\epsilon=-\int_D\frac{e^{ik[|x-y|-\beta\cdot
(x-y)]}}{4\pi|x-y|}\psi(y)dy, \qquad \psi:=qv. $$  
Using the Fourier transform  of convolution, one gets
\be\label{e13}
\tilde{\epsilon}=-F\Big(\frac{e^{ik[|x|-\beta\cdot x]}}{4\pi|x|}\Big)
F(qv), \qquad F(\psi):=\tilde{\psi}.
\ee

The assumption $q\in W^{\ell, 1}_0(D)$ and the elliptic regularity
results for $v$, which solves a second-order elliptic equation, 
imply that $v$ is smoother than $q$, and,
therefore,  $\psi=qv$
belongs to $W^{\ell,1}_0(D)$,
$\psi\in W^{\ell,1}_0(\mathbb{R}^3)$, $\ell>2$.

Let us now derive the estimate  \eqref{e14}, given below. 

If a function $f\in L^1(\R^3)$, then $|\tilde{f}|\leq c$. Here and below 
by  $c>0$ we denote various constants.

If $f\in W^{\ell,1}_0(D)$, then
$D^\ell f\in L^1(\R^3)$, where $D^\ell$ stands for any derivative 
of order $\ell$.
Therefore $|F(D^\ell f)|=|\xi^\ell \tilde{f}|\leq c$. If $f$ is compactly supported, then
$ \tilde{f}\in C^\infty_{loc}(\R^3)$, and the estimate $|\xi^\ell \tilde{f}|\leq c$
implies the inequality 
$$\sup_{\xi\in \R^3}(1+|\xi|)^\ell |\tilde{f}|<c.$$
We apply this inequality to the function $f=qv:=\psi\in  W^{\ell,1}_0(D)$ and get:
\be\label{e14}
(1+|\xi|)^\ell |\tilde{\psi}|<c, \qquad \ell>2.
\ee

Let us calculate now the first factor on the right-hand side of equation  \eqref{e13}. We have 
\be\label{e15}
\int_{\R^3}e^{i\xi \cdot x} \frac {e^{ik[|x|-\beta \cdot x]}}{4\pi |x|}=
-\frac 1 {|\xi|^2 -2k\beta \cdot \xi}.
\ee
Therefore
\be\label{e16}
\tilde{\epsilon}=-\frac {\tilde{\psi}(\xi)}{|\xi|^2 -2k\beta \cdot \xi}.
\ee

Let us replace $k$ by $k+i\eta$ in  \eqref{e15} and \eqref{e16}. 
In $\tilde{\psi}$ the dependence on $k$ enters through $v$.  
 Choose $\eta>\eta_0>0$  sufficiently large, so that the integral $I$ in  
\eqref{e18} (see below) will be as small as we wish.
This will yield estimate \eqref{e11} with $k+i\eta$ replacing $k$.

Using the spherical coordinates with the $z-$axis directed along $\beta$, $t=\cos \theta$,
$\theta$ is the angle between $\beta$ and $x-y$, $r:=|x-y|$, and using
 estimate \eqref{e14}, one gets:
\be\label{e17}
||\tilde{\epsilon}||_1\leq c\int_0^\infty \frac {dr 
r}{(1+r)^{\ell}}\int_{-1}^{1}    \frac{dt}{[|r-2kt|^2+4\eta^2 t^2]^{1/2}}:=cI.
\ee
The integral with respect to $t$ in  \eqref{e17} can be calculated in
closed form, and one gets:
\be\label{e18}
I=\frac 1 {2(k^2+\eta^2)^{1/2}}\int_0^\infty \frac {dr r}{(1+r)^\ell}\log\Big 
|\frac {1-a+[(1-a)^2+b]^{1/2}}{-1-a+[(1+a)^2+b]^{1/2}}\Big |,
\ee
where 
\be\label{e19}
a:=\frac {kr}{2(k^2+\eta^2)},\qquad b:=\frac {\eta^2r^2}{4(k^2+\eta^2)}.
\ee
If $r\to \infty$, then the ratio under the log sign in  \eqref{e18}
tends to $1$, and, since $\ell>2$,  the integral in 
\eqref{e18} converges. 

If $\eta>0$ is sufficiently large, then estimate \eqref{e18} implies that the 
inequality 
\eqref{e11} holds with $k$ replaced by $k+i\eta$. Therefore 
$\tilde{p}(2(k+i\eta)\beta)=0$ $\forall k>0, \, \forall \beta\in S^2$ and $\eta>\eta_0$.
This implies $\tilde{p}=0$, so $p=0$, and the first claim 
of Theorem 1 is proved.

The second claim of Theorem 1 is proved similarly. One starts with the orthogonality relation
$$\int_D p(x)u_1(x,\alpha_0, k) u_2(x, \beta, k)dx=0\quad \forall k>0, \,  \forall \beta\in 
S^2,$$  
writes it as
$$\int_D p(x)e^{ik(\alpha_0+\beta)\cdot x}[1+\epsilon]dx=0\quad \forall k>0, \,  \forall \beta\in
S^2,$$
and, replacing $k$ with $k+i\eta$, gets
$$\tilde{p}((k+i\eta)(\alpha_0+\beta))+(2\pi)^{-3}\tilde{p}\star \tilde{\epsilon}=0.$$
Using estimate  \eqref{e11} with $k+i\eta$ replacing $k$, one obtains the relation 
$$\tilde{p}((k+i\eta)(\alpha_0+\beta))=0\quad \forall k>0, \,  \forall \beta\in
S^2,\quad \eta>\eta_0.$$ Since $\tilde{p}(\xi)$ is an entire function of $\xi\in \C^3$,
this implies $\tilde{p}=0$, so $p=0$, and the second claim of Theorem 1 is proved.

Theorem 1 is proved \hfill $\Box$

\end{document}